\documentclass[11pt]{article}
\usepackage{epsfig}
\usepackage{amssymb}
\usepackage{doublespace}

\def\etal{{\it et al.}}
\def\Journal#1#2#3#4{{#1} {\bf #2}, #3 (#4)}

\def\AA{\em A.\& A.}
\def\APJ{\em ApJ.}
\def\APP{\em Astropart. Phys.}
\def\CPC{\em Comp. Phys. Com.}
\def\EJP{{\em Europ. J. Phys.} C}
\def\GEN{}
\def\IMP{\em Int. J. Mod. Phys.}
\def\IMP{\em Int. J. Mod. Phys.}
\def\JHE{\em J. High Ener. Phys.}
\def\JPG{\em J. Phys. G: Nucl. Part. Phys.}

\def\JPL{\em JETPhys. Lett.}
\def\JPT{\em Sov. Phys. JETP}
\def\MPA{{\em Mod. Phys.} A}

\def\NAT{\em Nature}

\def\NPB{{\em Nucl. Phys.} B}
\def\PLB{{\em Phys. Lett.}  B}
\def\PRL{\em Phys. Rev. Lett.}
\def\PRD{{\em Phys. Rev.} D}
\def\PRV{\em Phys. Rev.}
\def\PRE{\em Phys. Rep.}
\def\SCI{\em Science}
\def\SSR{\em Space Sci. Rev.}

\def\be{\begin{equation}}
\def\ee{\end{equation}}
\def\bea{\begin{eqnarray}}
\def\eea{\end{eqnarray}}

\begin{document}
\begin{spacing}{1.5}

\begin{center}
\Large \bf { Searching the Footprint of WIMPZILLAs \\}
\end{center} 

\begin{center}
{\it Houri Ziaeepour\\{ESO, Schwarzchildstrasse 2, 85748, Garching b. 
M\"{u}nchen, Germany\footnote{Present Address: 03, impasse de la Grande 
Boucherie,
F-67000, Strasbourg, France.}\\
Email: {\tt houri@eso.org}}}
\end{center}

\begin {abstract}
We constrain mass, lifetime and contribution of a very slowly decaying Ultra 
Heavy Dark Matter (UHDM) by simulating the cosmological evolution of its 
remnants. Most of interactions which participate in energy dissipation are 
included in the numerical solution of the Boltzmann equation. Cross-sections 
are calculated either analytically or by using PYTHIA Monte Carlo program. 
This paper describes in detail our simulation. To show the importance of the 
distribution of matter in constraining WIMPZILLA~\cite {wimpzilla} 
characteristics, 
we consider two extreme cases: a homogeneous universe, and a local halo with 
uniform distribution. In a homogeneous universe, the decay of the 
UHDM with a mass $\sim 10^{15} GeV$ and a lifetime as short as a 
few times of the age of the Universe, can not explain the flux of the 
observed Ultra High Energy Cosmic Rays (UHECRs) even if the whole Dark Matter 
(DM) is composed of a decaying UHDM. If our simple halo model is 
reliable, in a uniform clump with an over-density of $\sim 200$ 
extended to $\sim 100 kpc$, the lifetime can be $\sim 10 - 100 \tau_0$, again 
assuming that DM is a decaying UHDM. We also compare our calculation 
with observed $\gamma$-rays at $E \sim 10^{11} eV$ by EGRET and CASA-MIA limit 
at $E \sim 10^{15} eV$. They are compatible with a UHDM with relatively short 
lifetime.\\

\end {abstract}
\pagebreak
\section {First Encounter}
Particle Physics today is a zoo with plenty of wild species very difficult to 
trace or capture. They are in general known under phyla WIMPs, LSP, Axions, 
Higgs, Heavy Neutrinos, etc. Most of these species are potential candidates 
of the Dark Matter.\\
Recently a new phylum called WIMPZILLA~\cite{grapro} has been added to this 
zoo. The only common characteristic of the members of this family is their 
enormous mass, close to GUT scale $10^{16} GeV$ and their presumed long 
lifetime, much larger than the age of the Universe.\\
Theoretical motivations for existence of these particles is not very 
new. Since early 90s, some compactification scenarios in string theory have 
predicted composite particles (e.g {\it cryptons}) with large symmetry 
groups~\cite {crypton1} and $M \gtrsim 10^{14} GeV$. M-theory~\cite{mth} 
provides better candidates if compactification scale is much larger than 
Standard Model weak interaction scale~\cite {crypton2}. Messenger bosons 
in soft supersymmetry breaking models (see ~\cite{susy} for 
review) also can have close to GUT scale masses. Being composite and decaying 
only through non-renormalizable interactions or having discrete gauge 
symmetry~\cite{dissy} can make these ultra heavy particles meta-stable. 
Parametric resonance~\cite{reso} or vacuum fluctuation~\cite{grapro} at the 
end of inflation can produce a large amount of UHDM and unitarity constraint 
on their mass~\cite{uni} can be overcome if they have never been 
thermalized~\cite{grapro}.\\
If there was not other motivation for existence of an ultra heavy meta-stable 
particle, it was just one of many predictions of Particle Physics models  
waiting for detection. However, the {\it discovery} of Ultra High Energy 
Cosmic Rays (UHECRs) by large Air Shower detectors~\cite{yakutsk} 
~\cite{fly} ~\cite{agasa} (For a review of UHECRs detection and observed 
properties see ~\cite{crrev}) is an observational motivation. 
The predicted GZK cutoff~\cite{cmbir} in the spectrum of CRs at energies 
around $\sim 10^{18} eV$ due to interaction with CMBR and IR photons 
restricts the distance to sources to less than 
$20-30 Mpc$. At present there are about $600$ events with $E \sim 10^{19} eV$, 
$50$ with $E > 4 \times 10^{19} eV$ and 14 events with 
$E > 10^{20} eV$~\cite{eveclust}~\cite{evelist} including one with 
$E \sim 10^{21} eV$~\cite{fly}. The UHECR spectrum has a local minimum around 
$E \sim 10^{19} eV$ but it rises again at higher energies.\\
Composition of the primary particles~\cite {compo}~\cite {comp2} can be 
estimated from the shower and its muon content maximum position and their  
elongation rate in the atmosphere. In spite of uncertainties and 
dependence on hadrons interaction models at high energies~\cite{hadmod}, 
all analyses of the data are compatible with a composition change from iron 
nuclei to proton at 
$E > 10^{18} eV$~\cite{fly}~\cite{compo}~\cite{comp2}~\cite {nuclabs}. 
Based on theoretical arguments however, some authors suggest that 
events with highest energies can be produced by heavy nuclei~\cite{nuclei} (But see 
also ~\cite {nuclabs} for maximum fly distance of $Fe$ nuclei).\\
Some of ideas about the origin of UHECRs has been reviewed 
in ~\cite {revorg} and references therein. For the sake of 
completeness here we briefly review arguments for and against conventional 
and exotic sources.\\
{\bf Classical Candidates:} In a recent review of conventional 
candidates of UHECR sources~\cite{stdsrc}, R. Blandford rules out practically 
all of them.\\
It is expected that charged particles are accelerated in the 
shock waves of AGNs, SNs, or in-falling gas in rich galaxy clusters. 
Their maximum energy is proportional to the magnetic field strength and the 
size of the accelerator. Using this simple estimation, remnants of 
old SNs are expected to accelerate protons up to $\sim 10^{15} eV$ and iron to 
$\sim 10^{18} eV$~\cite{src1}. A $\sim 100 TeV$ $\gamma$-ray is expected from 
synchrotron radiation of these protons but nothing has been 
observed~\cite{snhigh}. Shock front of AGNs relativistic jets can 
accelerate protons to $E \sim 10^{20} eV$~\cite{accagn}. Central black 
hole in Fannaroff-Riley galaxies ~\cite {fanngal} or the remnant of 
QSO~\cite{qsorem} can produce UHECRs up to $E \sim 10^{20} eV$ with a 
marginal maximum energy of 
$4 \times 10^{21} eV$~\cite{maxacc}. Adiabatic expansion~\cite{exnu} and 
{\it in situ} interaction with radiation and matter fields however reduce 
the achievable maximum energy by orders of magnitude. Abrupt termination of 
the acceleration zone or composition change from $p$ to $n$~\cite{exnu} 
increases the chance of particles to keep their energies. The former however 
needs a fine tuning of the source structure and in the latter case 
particles lose part of their energy anyway.\\
Gamma Ray Bursts (GBR) also have been proposed as a candidate source of 
UHECRs~\cite{grb}. A simulation~\cite{gphdel} of cosmological distribution 
of sources with a power law flux of UHECRs shows however that the expected 
flux on Earth is much lower than observed value.\\
In ~\cite {crprog3} it is argued that Poisson noise in the process 
of proton interaction with background photons leaves a non-interacting tail in 
the flux of UHECRs and increases the probability of detecting UHECRs from 
further distances. Optical depth of protons around GZK cutoff can be roughly 
estimated by $\tau_{opt} \approx \sigma n_{cmb}$. For $\sigma \approx 0.45 
mb$ close to the resonance, the probability of non-interacting in a 
distance of $30 Mpc$ is at most $\sim 10^{-8}$.\\
{\bf Correlation with Astronomical Objects:} At present no serious 
astronomical candidate source has been observed. A correlation 
between Super Galactic Plane and UHECR events direction 
has been claimed~\cite{corr}, but ruled out by other analyses~\cite{nocorr}. 
It is plausible that the apparent clustering of events reported by AGASA 
collaboration~\cite {eveclust} originates from caustics generated by the 
galactic magnetic field~\cite {crprog1}~\cite {crprog2}~\cite {magdev}~\cite {virgo}.\\
Most of UHECRs burst models predict only one important nearby 
source~\cite{grb}~\cite{crprog3}. M87 in Virgo Cluster is practically the only 
conventional candidate that respects the distance constraint and is probably 
able to accelerate protons to ultra high energies. Recently, it has been shown 
that galactic wind and its induced magnetic field can deflect protons with 
$E > 10^{20} eV$ if the magnetic field is as high as $7 \mu G$. In this case, 
most of UHECR events point to Virgo Cluster~\cite {virgo}. Even with 
such strong magnetic field, this model can correlate most energetic 
events with M87/Virgo only if the primaries are $He$ nuclei.\\
There are also claims of correlation between the direction of UHECRs and 
pulsars~\cite {pulsar} or QSOs~\cite {qso}. The latter is reliable only if ultra high energy $\nu$s produce protons by interacting with relic or 
a halo of $\nu$~\cite {neuthalo}. Recent works~\cite {againtnu} however rule 
out a large part of the parameter space.\\
Close to uniform distribution of UHECRs events has been 
concluded to be the evidence that UHECRs originate from some 
extra-galactic astronomical objects and not from a decaying UHDM in the 
Galactic Halo~\cite{hanti}~\cite{hanti1}~\cite{18ev}. 
MACHOs observation~\cite{machobs} however shows that the Halo has a 
heterogeneous composition and a precise modeling of the anisotropy must take 
into account the distribution of various components.\\
{\bf Exotic Sources:} Ultra Heavy particles can be either long life DM or 
short life particles produced by the decay of topological 
defects~\cite{xpart}. In the latter case, the heavy particles 
decay in their turn to ordinary species and make the observed UHECRs. The 
result of the decay of a heavy short life particle and a heavy relic can 
be quite the same, but their production rate and its cosmological evolution 
are very different and depend on the defect type and model~\cite {defect}.  
The possibility that topological defects be the source of UHECRs has been 
studied extensively~\cite {defect}~\cite {crpro}~\cite {revorg}. Nonetheless, 
the observed power spectrum of LSS and CMB anisotropies rule out the 
existence of large amount of defects in the early 
Universe~\cite{againstdef}~\cite{inflpstr} and consequently the chance that 
UHECRs be produced by their decay (there are also limits from high energy 
$\nu$~\cite {revorg}).\\
Another proposed source of UHECRs is the evaporation of primordial black 
holes (PBH). The Hawking temperature at the end of their life is enough 
high to produce extremely energetic elementary particles like quarks and 
gluons and thus UHECRs. Most of models for production 
of PBH however needs fine tuning. Moreover, they are mainly produced when a 
large over density region crosses the horizon~\cite {pbh}. PBHs with present 
temperature of the same order as UHECRs energy must have an initial mass of 
$10^{14}-10^{15}gr$ and thus formed when the temperature of the Universe was  
$\sim 10^9 GeV$. A thermal inflation~\cite {pgheva} at EW scale would have 
reduced their density $\sim 10^{12}$ times. In models with low reheating 
temperature, at these scales the Universe is not yet thermalized and 
parametric resonance and fluctuation production happens only at superhorizon 
scales~\cite {reso}.\\
Summarizing the discussion of this section, it seems that conventional sources 
of Cosmic Rays are not able to explain the observed rate of 
UHECRs~\cite {stdsrc}. Between exotic sources, the decay of a meta-stable 
ultra heavy particle seems to be the most plausible one.

\section {WIMPZILLA Decay and Energy Dissipation of Remnants}
The decay of UHDM can have important implications for the evolution of high 
energy backgrounds. This can also be used for verifying this 
hypothesis and constraining the mass and lifetime of these particles. A number 
of authors have already tried to estimate the possible range of parameters 
as well as the flux of remnants on Earth (see ~\cite{defcal} and~\cite{crpro} 
for defects, and ~\cite {xx},~\cite{sarkar},~\cite {houri} and~\cite{topdown} 
for UHDM). 
In the present work more interactions have been included in the 
energy dissipation of UHDM remnants. Hadronization is implemented using 
PYTHIA Monte Carlo program~\cite{pythia}. We 
consider two distributions for UHDM. First we study the evolution of the 
spectrum of stable particles i.e. $e^{\pm}, p^{\pm}, \nu, \bar {\nu}$ and 
$\gamma$ in a homogeneous universe from photon decoupling to today. Then to 
show the effect of matter clumpiness, we simulate the Galactic Halo by 
simply considering a uniform over-density. A complete treatment of a clumpy 
universe will be reported elsewhere. We also estimate the effect of 
very slow decay of DM on the equation of state of the Universe and on the 
baryon and lepton asymmetries.\\
In this section we describe the decay model of UHDM and interactions which 
are included in the simulation.

\subsection {Decay Model}
\begin{figure}[t]
\begin{center}
\psfig{figure=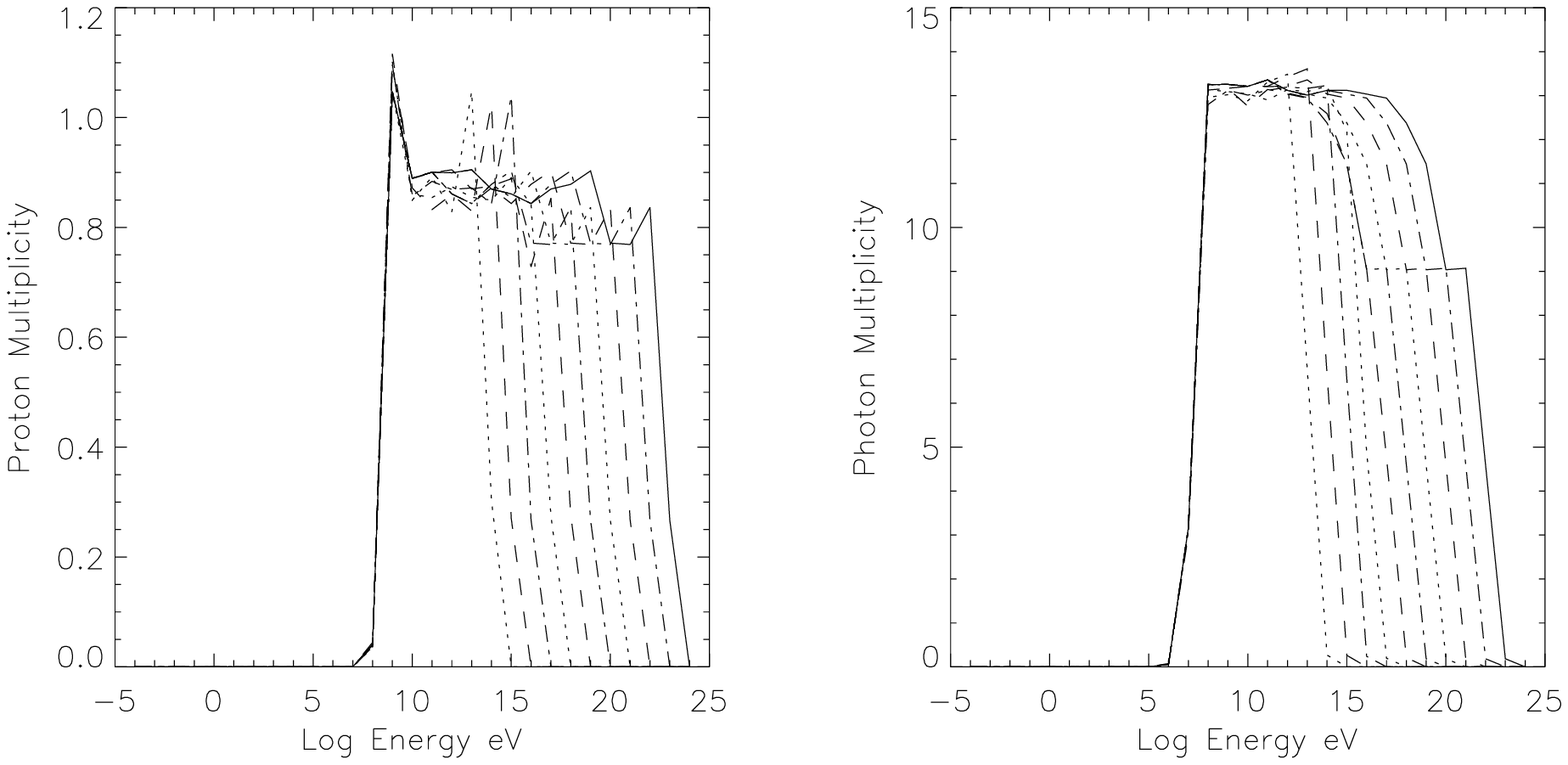,height=5cm}
\caption{Proton and photon multiplicity in hadronization of a pair of gluon 
jets for $E_{CM} = 10^{14}-10^{24} eV$. For $E_{CM} > 10^{20} eV$ the curves 
correspond to extrapolation from lower energies. 
\label{fig:decay}}
\end{center}
\end{figure}

Theoretical predictions for mass and lifetime of UH particles cover a large 
range of values $m_{UH} = 10^{22}-10^{26} eV$ and $\tau_{UH} = 
10^7-10^{20} yr$~\cite{crypton1}~\cite{crypton2}~\cite{highmass}. 
Nevertheless, at the end of inflation it is more difficult to produce 
the highest range of the masses and special types of inflationary models~\cite{instinf} are needed. For UH particles 
make a substantial part of the Dark Matter today, their lifetime must be at 
least comparable to the age of the Universe. In this work we perform the 
simulation for $m_{UH} = 10^{22} eV$ and $m_{UH} = 10^{24} eV$. For 
lifetime, $\tau_{UH} = 5 \tau_0$ and $\tau_{UH} = 50 \tau_0$, where $\tau_0$ 
is the age of the Universe, are studied. These values are smaller than what 
have been used 
by other groups~\cite{xx}~\cite{sarkar}~\cite{topdown}. We show below that taking into 
account a realistic model for energy dissipation of remnants, even these 
relatively short lifetime {\bf can not} explain the flux of UHECRs in a 
homogeneous universe. For some halo and IR background models, these lifetimes 
or slightly larger ones are compatible with observations.\\
The decay modes of UHDM are very model dependent. It is very likely that they 
don't decay directly to known particles and their decay has a number of 
intermediate unstable states that decay in their turn. It is also very 
probable 
that remnants include stable WIMPs which are not easily observable. To study 
the maximal effects of the decay on high energy backgrounds, we 
assume that at the end, the whole decayed energy goes to stable 
{\it visible} particles.\\
Most of WIMPZILLA models consider them to be neutral bosons. Due to lack 
of precise information about their decay, we assume that it looks like the 
decay of $Z^{\circ}$. Theoretical and experimental arguments show that 
leptonic and hadronic decay channels of $Z^{\circ}$ have a branching ratio 
of $\sim 1/3-2/3$~\cite{pphrev}. As hadronic channel is dominant, 
here we only consider this mode. It maximizes the flux of nucleons which at 
present are the dominant observable at ultra high energies.\\
To mimic the softening of energy spectrum due to multiple decay level, we 
assume that decay is similar to hadronization of a pair of gluon jets. 
Experimental data~\cite{qgdiff} as well as MLLA (Modified Leading 
Logarithm Approximation)~\cite {hadr}, LPHD (Local Parton-Hadron Duality)
(~\cite {lphd} and references therein) and string hadronization 
model~\cite {strhadr} predict a softer spectrum with higher multiplicity for 
gluon jets than for quarks.\\
We use PYTHIA program~\cite{pythia} for jet hadronization. This program, 
like many other available ones, can not properly simulate ultra high energy 
events, not only because we don't know the exact physics at $10^{16} GeV$ 
scale, but also because of programming limits. For this reason, we had to 
extrapolate simulation results for $E_{CM} \leqslant 10^{20} eV$ up to 
$E_{CM} = 10^{24} eV$. Fig.\ref {fig:decay} 
shows as an example, proton and photon multiplicity in hadronization of a pair 
of gluon jets. At middle energies, the multiplicity per 
$\log (E)$ is roughly constant. The same behavior exists for other 
species. This is a known shortcoming of present fragmentation 
simulations~\cite{fragrev} (See also Appendix 1) and makes spectrum harder 
at middle energies. As we would like to study the maximal flux of UHECRs and 
their effects on high energy backgrounds, this problem can not change our 
conclusions.\\
In the simulation, all particles except $e^{\pm}, p^{\pm}, \nu, \bar {\nu}$ 
and $\gamma$ decay. We neglect neutrinos mass and for simplicity we assume 
only one family of neutrinos i.e. $\nu_e$.\\
Contribution of the stable species in the total multiplicity and the total 
decay energy is 
summarized in Table \ref {tab:mult}. For all species, more than 
$99\%$ of the total energy belongs to the particles with energies higher than 
$10^{20} eV$ and $10^{18} eV$ respectively for two masses considered here. 
Apparently the mass of UHDM has little effect on the composition of remnants. 
However, one has to admit an uncertainty about this conclusion which is a 
direct consequence of the uncertain behavior of the multiplicity spectrum as 
mentioned above and in the Appendix 1.\\

\begin{table}[t]
\caption{Energy and multiplicity contribution in remnants of WIMPZILLA.
 \label{tab:mult}}
\vspace{0.2cm}
\begin{center}
\footnotesize
\begin{tabular}{|c|c|c|c|c|}
\hline
 & \multicolumn {2}{c|}{$M_{dm} = 10^{24} eV$} & 
\multicolumn {2}{c|}{$M_{dm} = 10^{22} eV$} \\
\hline
\raisebox{0pt}[13pt][7pt]{Part.} &
\raisebox{0pt}[13pt][7pt]{Ener. \%} &
\raisebox{0pt}[13pt][7pt]{Multi. \%} &
\raisebox{0pt}[13pt][7pt]{Ener. \%} &
\raisebox{0pt}[13pt][7pt]{Multi. \%}\\
\hline
\raisebox{0pt}[12pt][6pt]{$e^{\pm}$} & \raisebox{0pt}[12pt][6pt]{$6.7 \times 2$}
 & \raisebox{0pt}[12pt][6pt]{$9.7 \times 2$} & \raisebox{0pt}[12pt][6pt]{$6.7 \times 2$}
 & \raisebox{0pt}[12pt][6pt]{$9.8 \times 2$}\\
\raisebox{0pt}[12pt][6pt]{$p^{\pm}$} & \raisebox{0pt}[12pt][6pt]{$11.8 \times 2$}
 & \raisebox{0pt}[12pt][6pt]{$1.4 \times 2$} & \raisebox{0pt}[12pt][6pt]{$11.9 \times 2$}
 & \raisebox{0pt}[12pt][6pt]{$1.4 \times 2$}\\
\raisebox{0pt}[12pt][6pt]{${\nu} \& \bar{\nu}$} & \raisebox{0pt}[12pt]
[6pt]{$18.4 \times 2$} & \raisebox{0pt}[12pt][6pt]{$28.3 \times 2$}& \raisebox{0pt}[12pt][6pt]
{$18.1 \times 2$} & \raisebox{0pt}[12pt][6pt]{$28.3 \times 2$}\\
\raisebox{0pt}[12pt][6pt]{$\gamma$} & \raisebox{0pt}[12pt][6pt]{$26.2$}
 & \raisebox{0pt}[12pt][6pt]{$21$}& \raisebox{0pt}[12pt][6pt]{$26.6$}
 & \raisebox{0pt}[12pt][6pt]{$21$}\\
\hline
\end{tabular}
\end{center}
\end{table}
\subsection {Interactions}
\begin{figure}[t]
\begin{center}
\psfig{figure=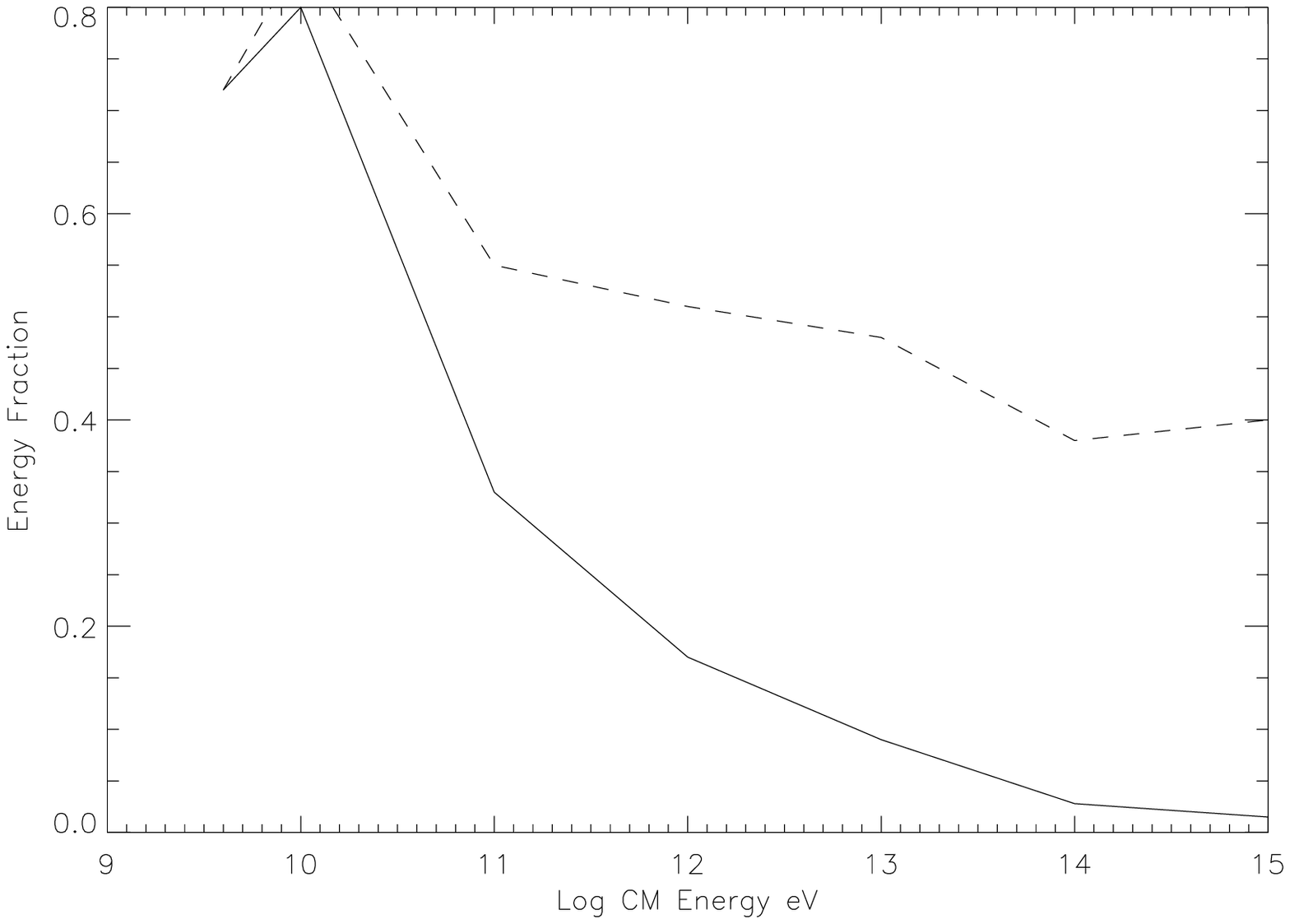,height=5cm}
\caption{Energy fraction of leading proton 
(dashed line) and mean energy of protons} 
{(solid line) with respect to the CM energy in $p-\gamma$ interaction.
\label{fig:pgmult}}
\end{center}
\end{figure}

We have included roughly all relevant interactions between 
remnants (except $\nu - \nu$ and $\bar \nu - \bar \nu$ elastic scattering) to the simulation either 
analytically or by using the results of PYTHIA Monte Carlo. Previous 
works~\cite{defcal}~\cite{xx}~\cite{crpro}~\cite{sarkar}~\cite{topdown} either don't 
consider the energy dissipation or take into account only the first order 
perturbative interactions (except for $p-\gamma$ where a fitting is used for 
$N-\gamma \rightarrow N-\pi$ cross-section). An early version of the 
present simulation~\cite {houri} studied only energy dissipation of UHECRs 
using PYTHIA 
without considering low energy processes and interactions of secondary 
particles. It found a higher lifetime for UHECRs than present work.\\
The main reason for considering only first order interactions is that it is 
usually assumed that interaction with CMB is dominated by the minimal process 
i.e $N-\gamma \rightarrow N-\pi$. However, the CMB spectrum even on its peak 
spans over a wide range of energies where radiation correction and 
hadronization become important. For instance, at $E_{CM} = 4 GeV$ the mean 
multiplicity is $\sim 15$ in place of 5 (after pion decay) in the minimal 
interaction. Moreover, 
in the galactic medium, the IR and visible radiations are comparable with CMB 
and play an important role in the energy dissipation of protons of $E \sim 
10^{18} - 10^{19}eV$. In extragalactic medium, the number density of 
background high energy photons with ($E > 1 eV$) is larger than visible and 
near IR. Another factor which accelerates energy dissipation of protons in the 
interaction with high energy photons is energy loss of leading proton in 
$p-\gamma$ interaction. It increases with energy (See Fig.\ref{fig:pgmult}) 
and results a higher dissipation rate.\\
PYTHIA can not simulate processes with invariant CM energies 
$E_{CM} \equiv \sqrt s$, $E_{CM} < 2 GeV$ to $E_{CM} < 4 GeV$ (the lower limit 
depends on the interaction). Consequently, 
for smaller energies we have included only perturbative, first order 
interactions using analytical expressions. Table \ref{tab:interact} summarizes 
processes which are included in the simulation, the energy range of 
analytical and/or Monte Carlo calculation of the cross-sections, and the cuts 
used for removing singularities at small energies or angles in the case of 
analytical calculation (results depend somehow on these cuts specially in the 
case of moderate energy and angular resolution of our program). For energy 
ranges that PYTHIA has been used, in general we use 
default value of various parameters of the program as defined 
in PYTHIA manual. A number of parameters have been changed for all processes 
e.g. to make unstable particles like mesons decay etc. They are listed in 
Table \ref{tab:genpythia}. For some interactions, the value of a few other 
parameters are changed. They are also listed in the Table \ref{tab:interact}. 
From now on $s$, $t$ and $u$ are Mandelstam variables.\\
\begin{table}[t]
\caption{PYTHIA Parameters \label{tab:genpythia}}
\vspace{0.2cm}
\begin{center}
\footnotesize
\begin{tabular}{|c|l|}
\hline
\raisebox{0pt}[12pt][6pt] {Selected processes} & \raisebox{0pt}[12pt][6pt]
{lept.-lept, lept.-had., unresolved $\gamma$, low $p_t$, diffract. } \\
\raisebox{0pt}[12pt][6pt] & \raisebox{0pt}[12pt][6pt]
{\& elastic had.-had., Radiative correction}\\
\hline
\raisebox{0pt}[12pt][6pt] {Status parameters} & \raisebox{0pt}[12pt][6pt]
{$2^{ed}$-order $\alpha_s$, continuous $p_t$ cutoff.}\\
\hline
\raisebox{0pt}[12pt][6pt] {Pt cuts} & 
\raisebox{0pt}[12pt][6pt] {$E_{CM}^{min} = 1$, $p_{t}^{min} = 0$}\\
\hline
\raisebox{0pt}[12pt][6pt] {Number of flavors} & \raisebox{0pt}[12pt][6pt]
{$6$}\\
\hline
\end{tabular}
\end{center}
\end{table}
\begin{table}[t]
\caption{Interactions \label{tab:interact}}
\vspace{0.2cm}
\begin{center}
\footnotesize
\begin{tabular}{|c|c|c|c|c|c|}
\hline
\raisebox{0pt}[13pt][7pt]{Interaction} &
\raisebox{0pt}[13pt][7pt]{Analyt. Cal.} &
\raisebox{0pt}[13pt][7pt]{Cuts} &
\raisebox{0pt}[13pt][7pt]{PYTHIA $E_{CM}$} &
\raisebox{0pt}[13pt][7pt]{PYTHIA Parameters} &
\raisebox{0pt}[13pt][7pt]{Ref. Analytic}\\
\raisebox{0pt}[13pt][7pt]{} &
\raisebox{0pt}[13pt][7pt]{$E_{CM}$ Rang} &
\raisebox{0pt}[13pt][7pt]{} &
\raisebox{0pt}[13pt][7pt]{Rang} &
\raisebox{0pt}[13pt][7pt]{} &
\raisebox{0pt}[13pt][7pt]{Cross-Sec.}\\
\hline
\raisebox{0pt}[12pt][6pt]{$e^\pm - e^\pm$ elastic} & 
\raisebox{0pt}[12pt][6pt]{All}
 & \raisebox{0pt}[12pt][6pt] {$\frac {s - 4m_e^2}{m_e^2} > 10^{-3}$,} & 
 \raisebox{0pt}[12pt][6pt]{-}
 & \raisebox{0pt}[12pt][6pt]{-} & \raisebox{0pt}[12pt][6pt]{~\cite{landau}}\\
\raisebox{0pt}[12pt][6pt]{} & 
\raisebox{0pt}[12pt][6pt]{}
 & \raisebox{0pt}[12pt][6pt]{$|\frac {t}{s}| > 10^{-2}$,$|\frac {u}{s}| > 10^{-2}$} & 
 \raisebox{0pt}[12pt][6pt]{}
 & \raisebox{0pt}[12pt][6pt]{} & \raisebox{0pt}[12pt][6pt]{}\\
\hline
\raisebox{0pt}[12pt][6pt]{$e^+ - e^-$ elastic} & 
\raisebox{0pt}[12pt][6pt]{$E_{CM} \leq 4 GeV$}
 & \raisebox{0pt}[12pt][6pt]{"} & 
 \raisebox{0pt}[12pt][6pt]{included in $e^+ - e^-$}
 & \raisebox{0pt}[12pt][6pt]{-} & \raisebox{0pt}[12pt][6pt]{"}\\
\raisebox{0pt}[12pt][6pt]{} & 
\raisebox{0pt}[12pt][6pt]{}
 & \raisebox{0pt}[12pt][6pt]{} & 
 \raisebox{0pt}[12pt][6pt]{$\rightarrow\ldots$}
 & \raisebox{0pt}[12pt][6pt]{-} & \raisebox{0pt}[12pt][6pt]{"}\\
\hline
\raisebox{0pt}[12pt][6pt]{$e^+ - e^- \rightarrow 2\gamma$} & 
\raisebox{0pt}[12pt][6pt]{"}
 & \raisebox{0pt}[12pt][6pt]{"} & 
 \raisebox{0pt}[12pt][6pt]{"}
 & \raisebox{0pt}[12pt][6pt]{-} & \raisebox{0pt}[12pt][6pt]{"}\\
\hline
\raisebox{0pt}[12pt][6pt]{$e^+ - e^- \rightarrow \ldots$} & 
\raisebox{0pt}[12pt][6pt]{-}
 & \raisebox{0pt}[12pt][6pt]{-} & 
 \raisebox{0pt}[12pt][6pt]{$4 GeV \leq E_{CM} \leq 10^{6}$ }
 & \raisebox{0pt}[12pt][6pt]{Init. Braamst.} & \raisebox{0pt}[12pt][6pt]{-}\\
\hline
\raisebox{0pt}[12pt][6pt]{$p^\pm - e^\pm \rightarrow \ldots$} & 
\raisebox{0pt}[12pt][6pt]{-} & \raisebox{0pt}[12pt][6pt]{-} & 
 \raisebox{0pt}[12pt][6pt]{"}
 & \raisebox{0pt}[12pt][6pt]{-} & \raisebox{0pt}[12pt][6pt]{-}\\
\raisebox{0pt}[12pt][6pt]{(all combinations)} & 
\raisebox{0pt}[12pt][6pt]{} & \raisebox{0pt}[12pt][6pt]{} & 
 \raisebox{0pt}[12pt][6pt]{}
 & \raisebox{0pt}[12pt][6pt]{} & \raisebox{0pt}[12pt][6pt]{}\\
\hline
\raisebox{0pt}[12pt][6pt]{$p^\pm - p^\pm \rightarrow \ldots$} & 
\raisebox{0pt}[12pt][6pt]{-} & \raisebox{0pt}[12pt][6pt]{-} & 
 \raisebox{0pt}[12pt][6pt]{$3 GeV \leq E_{CM} \leq 10^{5}$}
 & \raisebox{0pt}[12pt][6pt]{-} & \raisebox{0pt}[12pt][6pt]{-}\\
\raisebox{0pt}[12pt][6pt]{(all combinations)} & 
\raisebox{0pt}[12pt][6pt]{} & \raisebox{0pt}[12pt][6pt]{} & 
 \raisebox{0pt}[12pt][6pt]{}
 & \raisebox{0pt}[12pt][6pt]{} & \raisebox{0pt}[12pt][6pt]{}\\
\hline
\raisebox{0pt}[12pt][6pt]{$\nu - e^-$ elastic} & 
\raisebox{0pt}[12pt][6pt]{All}
 & \raisebox{0pt}[12pt][6pt]{$\frac {s - m_e^2}{m_e^2} > 10^{-3}$} & 
 \raisebox{0pt}[12pt][6pt]{-}
 & \raisebox{0pt}[12pt][6pt]{-} & \raisebox{0pt}[12pt][6pt]{~\cite{neutcross}}\\
\hline
\raisebox{0pt}[12pt][6pt]{$\bar \nu - e^+$ elastic} & 
\raisebox{0pt}[12pt][6pt]{"} & \raisebox{0pt}[12pt][6pt]{"} & 
 \raisebox{0pt}[12pt][6pt]{-}
 & \raisebox{0pt}[12pt][6pt]{-} & \raisebox{0pt}[12pt][6pt]{"}\\
\hline
\raisebox{0pt}[12pt][6pt]{$\nu - e^+ \rightarrow \ldots $} & 
\raisebox{0pt}[12pt][6pt]{-}
 & \raisebox{0pt}[12pt][6pt]{-} & 
 \raisebox{0pt}[12pt][6pt]{$4 GeV \leq E_{CM} \leq 10^{6}$ }
 & \raisebox{0pt}[12pt][6pt]{-} & \raisebox{0pt}[12pt][6pt]{-}\\
\hline
\raisebox{0pt}[12pt][6pt]{$\bar \nu - e^- \rightarrow \ldots $} & 
\raisebox{0pt}[12pt][6pt]{-} & \raisebox{0pt}[12pt][6pt]{-} & 
 \raisebox{0pt}[12pt][6pt]{"}
 & \raisebox{0pt}[12pt][6pt]{-} & \raisebox{0pt}[12pt][6pt]{-}\\
\hline
\raisebox{0pt}[12pt][6pt]{$\nu - p^\pm \rightarrow \ldots $} & 
\raisebox{0pt}[12pt][6pt]{-}
 & \raisebox{0pt}[12pt][6pt]{-} & 
 \raisebox{0pt}[12pt][6pt]{$4 GeV \leq E_{CM} \leq 10^{6}$ }
 & \raisebox{0pt}[12pt][6pt]{-} & \raisebox{0pt}[12pt][6pt]{-}\\
\hline
\raisebox{0pt}[12pt][6pt]{$\bar \nu - p^\pm \rightarrow \ldots $} & 
\raisebox{0pt}[12pt][6pt]{-} & \raisebox{0pt}[12pt][6pt]{-} & 
 \raisebox{0pt}[12pt][6pt]{"}
 & \raisebox{0pt}[12pt][6pt]{-} & \raisebox{0pt}[12pt][6pt]{-}\\
\hline
\raisebox{0pt}[12pt][6pt]{$\nu - \bar \nu$} & 
\raisebox{0pt}[12pt][6pt]{-} & \raisebox{0pt}[12pt][6pt]{-} & 
 \raisebox{0pt}[12pt][6pt]{"}
 & \raisebox{0pt}[12pt][6pt]{-} & \raisebox{0pt}[12pt][6pt]{-}\\
\hline
\raisebox{0pt}[12pt][6pt]{$\gamma - e\pm \rightarrow \gamma - e\pm$} & 
\raisebox{0pt}[12pt][6pt]{$E_{CM} \leq 4 GeV$}
 & \raisebox{0pt}[12pt][6pt]{$\frac {s - m_e^2}{m_e^2} > 10^{-3}$,} & 
 \raisebox{0pt}[12pt][6pt]{-}
 & \raisebox{0pt}[12pt][6pt]{-} & \raisebox{0pt}[12pt][6pt]{~\cite{landau}}\\
\raisebox{0pt}[12pt][6pt]{} & 
\raisebox{0pt}[12pt][6pt]{}
 & \raisebox{0pt}[12pt][6pt]{$|\frac {t}{s}| > 10^{-2}$, $|\frac {u}{s}| > 10^{-2}$} & 
 \raisebox{0pt}[12pt][6pt]{}
 & \raisebox{0pt}[12pt][6pt]{} & \raisebox{0pt}[12pt][6pt]{}\\
\hline
\raisebox{0pt}[12pt][6pt]{$\gamma - e\pm \rightarrow \ldots$} & 
\raisebox{0pt}[12pt][6pt]{-}
 & \raisebox{0pt}[12pt][6pt]{-} & 
 \raisebox{0pt}[12pt][6pt]{$4 GeV \leq E_{CM} \leq 10^{6}$ }
 & \raisebox{0pt}[12pt][6pt]{-} & \raisebox{0pt}[12pt][6pt]{-}\\
\hline
\raisebox{0pt}[12pt][6pt]{$\gamma - p\pm \rightarrow \ldots$} & 
\raisebox{0pt}[12pt][6pt]{$1 GeV< E <$} & \raisebox{0pt}[12pt][6pt]{-} & 
 \raisebox{0pt}[12pt][6pt]{"}
 & \raisebox{0pt}[12pt][6pt]{${E_{CM}}^{min} = 1 GeV$,} & \raisebox{0pt}[12pt][6pt]{~\cite{pphrev}}\\
\raisebox{0pt}[12pt][6pt]{} & 
\raisebox{0pt}[12pt][6pt]{$4 GeV$} & \raisebox{0pt}[12pt][6pt]{} & 
 \raisebox{0pt}[12pt][6pt]{}
 & \raisebox{0pt}[12pt][6pt]{${p_t}^{min} = 0.5 GeV$,} & 
\raisebox{0pt}[12pt][6pt]{}\\
\raisebox{0pt}[12pt][6pt]{} & 
\raisebox{0pt}[12pt][6pt]{} & \raisebox{0pt}[12pt][6pt]{} & 
 \raisebox{0pt}[12pt][6pt]{}
 & \raisebox{0pt}[12pt][6pt]{Singul. Cut $= 0.25 GeV$} & 
\raisebox{0pt}[12pt][6pt]{}\\
\hline
\raisebox{0pt}[12pt][6pt]{$\gamma - \gamma \rightarrow e^+ - e^-$} & 
\raisebox{0pt}[12pt][6pt]{$E_{CM} \leq 3 GeV$}
 & \raisebox{0pt}[12pt][6pt]{$\frac {s - m_e^2}{m_e^2} > 10^{-3}$,} & 
 \raisebox{0pt}[12pt][6pt]{-}
 & \raisebox{0pt}[12pt][6pt]{-} & \raisebox{0pt}[12pt][6pt]{~\cite{landau}}\\
\raisebox{0pt}[12pt][6pt]{} & 
\raisebox{0pt}[12pt][6pt]{}
 & \raisebox{0pt}[12pt][6pt]{$|\frac {t}{s}| > 10^{-2}$, $|\frac {u}{s}| > 10^{-2}$} & 
 \raisebox{0pt}[12pt][6pt]{}
 & \raisebox{0pt}[12pt][6pt]{} & \raisebox{0pt}[12pt][6pt]{}\\
\hline
\raisebox{0pt}[12pt][6pt]{$\gamma - \gamma \rightarrow \ldots$} & 
\raisebox{0pt}[12pt][6pt]{-}
 & \raisebox{0pt}[12pt][6pt]{-} & 
 \raisebox{0pt}[12pt][6pt]{$3 GeV \leq E_{CM} \leq 10^{6}$}
 & \raisebox{0pt}[12pt][6pt]{-} & \raisebox{0pt}[12pt][6pt]{-}\\
\hline
\end{tabular}
\end{center}
\end{table}
At very high energies, $E_{CM} > 10^{14} eV$ for nucleon-nucleon and $E_{CM} > 
10^{15} eV$ for other interactions, PYTHIA becomes very slow and the number 
of rejected events increases rapidly. For these energies we perform a linear 
extrapolation from lower energies as explained below. 

\section {Evolution}
We assume that non-baryonic Dark Matter is totally composed of slowly decaying 
UH particles. From (\ref{idec}) and (\ref{jdec}) below it is evident that 
width and fraction of UHDM in DM are degenerate and in the evolution 
equations, decreasing contribution is equivalent to increasing lifetime.\\
Boltzmann equation for space-time and energy-momentum distribution of a 
particle $i$ is~\cite{ehler} (We use units with $c = \hbar = 1$):
\bea
p^{\mu}{\partial}_{\mu} f^{(i)}(x,p) - ({\Gamma}^{\mu}_{\nu\rho} p^{\nu} 
p^{\rho} - e_i F^{\mu}_{\nu} p^{\nu}) \frac {\partial f^{(i)}}{\partial 
p^{\mu}} & = & -({\mathcal A}(x,p) + {\mathcal B}(x,p)) f^{(i)}(x,p) + 
{\mathcal C}(x,p) + \nonumber \\
 & & {\mathcal D}(x,p) + {\mathcal E}(x,p). \label {bolt}
\eea
\bea
{\mathcal A}(x,p) & = & {\Gamma}_i m_i.\label {idec}\\
{\mathcal B}(x,p) &= & \sum_j \frac {1}{(2\pi)^3 g_i} \int d\bar p_j f^{(j)}(x,p_j) A (s){\sigma}_{ij}(s).\label {absint}\\
{\mathcal C}(x,p) & = & \sum_j {\Gamma}_j m_j \frac {1}{(2\pi)^3 g_i} \int d\bar p_j f^{(j)}(x,p_j)\frac {d{{\mathcal M}^{(i)}}_j}{d\bar p}.\label {jdec}\\
{\mathcal D}(x,p) & = & \sum_{j,k}\frac {1}{(2\pi)^6 g_i}\int d\bar p_j d\bar p_k f^{(j)}(x,p_j)f^{(k)}(x,p_k) A (s) \frac {d{\sigma}_{j+k \rightarrow i+\ldots}}{d\bar p}.
\label {proint}
\eea
$x$ and $p$ are coordinate and momentum 4-vectors; $f^{(i)}(x,p)$ is the 
distribution of species $i$; $m_i$, $e_i$ and $\Gamma_i$, are its mass,  
electric charge and width $= 1/\tau_i$, $\tau_i$ is the lifetime; 
${\sigma}_{ij}$ is the total interaction cross-section of species $i$ and 
species $j$ at a fixed $s$; $\frac {d{\sigma}_{j+k \rightarrow i+\ldots}}
{d\bar p} = \frac {(2\pi)^3 E d\sigma}{g_i p^2 dp d\Omega}$ is the Lorantz 
invariant differential cross-section of production of $i$ in the 
interaction of $j$ and $k$; $g_i$ is the number of internal degrees of freedom 
(e.g. spin, color); $d\bar p = \frac {d^3p}{E}$. We treat 
interactions classically, i.e. we 
consider only two-body interactions and we neglect the interference between 
outgoing particles. It is a good approximation when the plasma is not 
degenerate. It is assumed that cross-sections include summation 
over internal degrees of freedom like spin; $\frac {d{{\mathcal M}^{(i)}}_{j}}
{d\bar p}$ is the differential multiplicity of species $i$ in 
the decay of $j$; ${\Gamma}^{\mu}_{\nu\rho}$ is the connection; 
$F^{\mu}_{\nu}$ an external electromagnetic field; and finally 
${\mathcal E}(x,p)$ presents all other external sources. $A (s)$ is a 
kinematic factor~\cite {kinfact}:
\be
A (p_i,p_j) = ((p_i.p_j)^2 - m_i^2m_j^2)^{\frac {1}{2}} = 
\frac {1}{2} ((s - m_i^2 - m_j^2)^2 - 4 m_i^2m_j^2)^{\frac {1}{2}}.
\label {kin}\\
\ee
The quantity $A\sigma$ presents the probability of an interaction.\\
In a homogeneous universe $f(x,p) = f(t, |{\bf} p|)$ and in (\ref{bolt}) the 
term corresponding to interaction with external electromagnetic field is zero. 
Therefore, to have a consistent formalism for evolution of distribution of all 
species, we don't include the synchrotron radiation of high energy electrons 
in a magnetic field.\\
We only consider the evolution of stable particles and slowly decaying UHDM. 
The term (\ref{idec}) concerns only UHDM. In (\ref{jdec}), the only non-zero 
term in the sum is the decay of UHDM. We assume that stable species don't 
have any interaction with UHDM and corresponding interaction integrals in 
(\ref{absint}) and (\ref{proint}) are zero. Due to the very large mass of 
UHDM, its momentum is negligible and we can assume 
that in comoving frame it is at rest. This permits to use its number density 
$n_{dm}$ which is more convenient for numerical calculation.\\
In a homogeneous universe the metric in comoving frame is:
\be
ds^2 = dt^2 - a^2(t) \delta_{ij} dx^i dx^j.
\ee
and with respect to local Lorantz frame (\ref {bolt}) to (\ref {proint}) take the 
following form (in the following the species index indicates one of the stable 
species):
\bea
\frac{\partial f^{(i)}(t,p)}{\partial t} - \frac {\dot {a}}{a} p 
\frac {\partial f^{(i)}}{\partial p} & = & \frac {1}{E}(-{\mathcal B}(t,p) 
f^{(i)}(t,p) + {\mathcal C}(t,p) + {\mathcal D}(t,p)). \label {bolth}\\
{\mathcal B}(t,p) & = & \sum_j \frac {1}{(2\pi)^2 g_i} \int dp_j 
\frac {p_j^2}{E_j} f^{(j)}(t,p_j) \int d(\cos {\theta}_{ij}) A (s) 
{\sigma}_{ij}(s).\label {absinth}\\
{\mathcal C}(t,p) & = & \frac {E}{4\pi g_i p^2}{\Gamma}_{dm} n_{dm} 
\frac {d{\mathcal M}^{(i)}}{dp}. \label {jdech}\\
{\mathcal D}(t,p) & = & \sum_{j,k}\frac {1}{(2\pi)^5 g_i}\int dp_j dp_k \frac {p_j^2}{E_j} 
\frac {p_k^2}{E_k} f^{(j)}(t,p_j) f^{(k)}(t,p_k) \int d(\cos {\theta}_{jk}) 
d(\cos {\theta}_{ji}) \nonumber\\
 & & d\phi_i A (s) \frac {d{\sigma}_{j+k \rightarrow i+\ldots}}{d\bar p}.\label {prointh}\\
\frac {dn_{dm}}{dt} + \frac {3\dot {a}}{a} n_{dm} & = & -{\Gamma}_{dm} 
n_{dm} \label {idech}.
\eea
In (\ref{absinth}) and (\ref{prointh}) $s$ depends on the angle between species 
$j$ and $k$, and $j$ and $i$. Consequently, it is not possible to use 
cross-sections integrated over angular variables.\\
Evolution of $a (t)$ is ruled by Einstein equation:
\bea
\frac {\dot {a}^2}{a^2} & = & \frac {8\pi G}{3} T_{00} + \frac {\Lambda}{3}. \label {einsteq}\\
T^{00}(t) & = & \sum_i \frac {g_i}{2\pi^2} \int dp p^2E f^{(i)}(t,p). 
\label {t00}
\eea
In a homogeneous cosmology, $T^{00}$ in Local Lorantz frame is the same 
as comoving frame and $T^{ii}_{comov} = a^{-2} T^{ii}_{Loc.Lor.}$.\\
Equations (\ref{bolth}) and (\ref{einsteq}) determine the 
cosmological evolution of species. Due to interaction terms, even in a 
homogeneous universe, 
these equations are non-linear and coupled. It is not therefore possible to 
solve them analytically. 
Evolution equation for DM can be solved analytically for a short period of 
time. For other species if in 
(\ref{bolth}) we consider absorption and production integrals as $t$ and $p$ 
dependent coefficients of a linear partial differential equation, (\ref{bolth})
can be solved analytically~\cite{mathref}. Giving the value of $f^{(i)}(t,p)$,
 $n_{dm}(t)$ and $a (t)$, at time $t$, we can then determine $a (t + 
\Delta t)$, $n_{dm}(t + \Delta t)$, ${\mathcal B}(t,p)$, ${\mathcal C}(t,p)$ 
and ${\mathcal D}(t,p)$ for a short time interval $\Delta t$. $f^{(i)}(t + 
\Delta t,p)$ would be obtain from solution of partial differential 
equation (\ref{bolth}) using difference method. The solution of metric and 
distributions in one step of numerical calculation are the followings:
\bea
a (t + \Delta t) & = & a (t) \exp (\Delta t (\frac {8\pi G}{3} T_{00} + 
\frac {\Lambda}{3})^{\frac {1}{2}}).\label {asol}\\
n_{dm}(t + \Delta t) & = & n_{dm}(t) \frac {a^3(t_0)}{a^3(t)} 
\exp (-\frac {t-t_0}{\tau}).\label {dmsol}\\
f^{(i)}(t + \Delta t,p) & = & (f^{(i)}(t, p') + \Delta t ({\mathcal C}(t,p') + 
{\mathcal D}(t,p'))) \exp (-{\mathcal B}(t,p')\Delta t) \label {fsol}\\
p' & = & \frac {a (t + \Delta t) p}{a (t)}. \label {psol}
\eea
This prescription is more precise than a pure numerical calculation using 
e.g. difference method.

\section {Numerical Simulation \label {numcal}}
What makes numerical calculation of (\ref{asol}) to (\ref{fsol}) 
difficult is the extension on roughly $34$ orders of magnitude of energy 
from $10^{-9} eV$ (radio background) to $10^{24} eV$ (mass of UHDM) (from now 
on we call this energy range ${\mathcal R}_E$). Physical processes in this 
vast energy range have varieties of behavior, resonances, etc. Moreover, 
species have distributions which are orders of magnitude different from each 
others. In other term, these equations are very stiff. Semi-analytic method 
explained above helps to increase the precision of the numerical calculation. 
However, the 5-dimensional integration in (\ref{prointh}) is extremely time 
and memory consuming and it is impossible to determine it with the same 
precision.\\
In the following we describe in detail the numerical calculation of 
(\ref{asol}) to (\ref{fsol}), as well as cosmological model, initial 
conditions and backgrounds which have been used.

\subsection {Multiplicity and Cross-section}
For processes simulated by PYTHIA, we need to calculate total and differential 
cross-sections (see (\ref{absinth}) and (\ref{prointh})). The former is given 
by the program itself. To determine the latter, we divide ${\mathcal R}_E$ to 
logarithmic bins (one per order of magnitude) and classify particles according 
to their momentum. The angular distribution of 
produced particles with respect to the axis of incoming particles in CM also 
is divided linearly to 90 bins. At a given $s$, the cross section in each bin 
$\Delta \sigma_{ij} = \sigma_{tot} N_{ij} / N$. $N_{ij}$ is the number of 
particles of 
a given species in the bin $ij$. $N$ is the total number of simulated events.\\
The same procedure is used for determination of $\frac {d{\mathcal M}}{dp}$. 
In this case it is not necessary to consider the angular distribution 
because in the rest frame of WIMPZILLA the decay has a spherical symmetry.\\
Because PYTHIA can not cover the totality of the energy range, for high 
energy bins we use a linear extrapolation in $\log p$. The contribution of 
these energies i.e. $E_{CM} \gtrsim 10^{6} GeV$ on the evolution of species is 
nevertheless small because the density of concerning particles is very low. 
The reason for adding them is not to have an artificial cut in the 
calculation.\\
As mentioned above, for most processes including $p-\gamma$, PYTHIA can not 
simulate the interaction with $E_{CM} < 4 GeV$. $p-\gamma$ is the most 
important process for the energy dissipation of protons specially in this 
uncovered energy range where interaction with CMB photon is concentrated. 
In these energies we use directly the total cross-section obtained from 
experience~\cite{pphrev}. For differential cross-section, we extrapolate 
angular distribution from higher energies and normalize it to the exact total 
cross-section.

\subsection {Evolution Equations}
In (\ref{absinth}) and (\ref{prointh}), the integrals over angular degrees 
of freedom can be separated from energy integrals and they don't depend on any 
cosmological or DM parameter. It is therefore very convenient to calculate 
them separately.\\
We divide each $180^\circ$ interval to 
$9$ bins and use trapezoid method for integration. For the single integral in 
(\ref{absinth}) a better resolution with $90$ bins has been used. However, our 
tests show that even the moderate resolution of $9$ bins gives, up to a few 
percents, the same results as the more precise integration. This is a 
reassuring 
results and means that the triple integrals in production term also must be 
enough correct even with a low resolution.\\
Calculation of production term for first-order interactions is more 
complicate. They are processes of type $2 part. \rightarrow 2 part$. In the 
CM frame where all cross-sections are determined, analytically or numerically, 
the incoming and outgoing particles have the same momentum. This is equivalent 
to having a Delta function in the integrand. The numerical realization of this 
function specially with a moderate resolution is very difficult. 
Consequently, one has to analytically absorb this function into integrand. 
Details of the calculation can be found in Appendix 2.\\
Numerical solution of evolution equation itself needs much better energy 
resolution due to stiffness of distributions. The resolution must be at least 
comparable to smallest quantities. Our tests show that a 
division to 680 logarithmic bins of the energy range ${\mathcal R}_E$ (i.e 
$20$ bins per one order of magnitude) gives an acceptable compromise between 
precision and calculation time.\\
We divide the 
interval $z = [z_{dec} - 0.001]$ to 30 logarithmic bins and the last 
step is from $z = 0.001$ to $z = 0$. The program is written in $C^{++}$ 
language and is highly modulable. It can be requested from the author.\\
To test the precision of our numerical calculation, we have run the program 
without interaction terms. The error on total $T^{00}$ is $0.7\%$ and on 
non-baryonic DM is practically zero. For other species it is $5.5\%$ to 
$7.5\%$. Including interaction terms but not the decay of DM gives the same 
answer. This test is crucial for correct simulation of thermal equilibrium of 
the Universe.

\subsection {Cosmology Model and Initial Conditions}
We consider a flat universe with present value of parameters as the 
followings: 
$\Omega_M = 0.3$, $\Omega_{\Lambda} = 0.7$, $h = H_0 / 100 km \sec^{-1} 
Mpc^{-1} = 0.7$ and $\Omega_b = 0.02 h^{-2}$.\\
We fix 
the decoupling redshift at $z_{dec} = 1100$. The distribution of species at 
that time was thermal with a temperature $T_{dec} = T_{cmb} (z_{dec} + 1) = 
0.26 eV$, $T_{cmb} = 2.728 K$~\cite {cobetemp} for $e^\pm$, $p^\pm$ and 
$\gamma$ and $\frac {4}{11} T_{dec}$ for $\nu$ and $\bar \nu$.\\
In the same way, one can determine the temperature of Dark Matter
\footnote{One should consider $T_{dm}$ as an estimation of kinetic energy 
scale rather than a real temperature because if ultra heavy particles exist, 
they could never be thermalized.} at decoupling~\cite{earlyu}:
\be
T_{dm}(z_{dec}) = \frac {{g_{*s}^{dec}}^{\frac {2}{3}} T_{dec}^2}
{{g_{*s}^{dm}}^{\frac {2}{3}} T_{dm-dec}}
\ee
$T_{dm-dec}$ is the decoupling temperature of the Dark Matter. If we assume that 
$T_{dm-dec} \sim 10^{16} eV$, $T_{dm}(z_{dec}) < 10^{-18} eV$. 
Therefore, our approximation $T_{dm} = 0$ is quite justified.\\
We know the density of species at present. Their initial value at decoupling 
depends on the equation of state of the Universe. However, it is exactly 
what we want to calculate! Consequently, we have to determine their value at 
decoupling approximately by neglecting the effect of UHDM decay. As the 
lifetime of WIMPZILLA is assumed to be much longer than the age of the 
Universe, the present value of densities after evolution must stay very close 
to our initial assumption.\\
We {\it define} the initial densities as the followings:
\bea
n_\gamma = n_{COBE} (z_{dec} + 1)^3  \quad & & \quad n_\nu = n_{\bar\nu} 
= 3 \times \frac {4}{11} \times n_\gamma\\
n_p = \frac {\Omega_b \rho_c}{m_p} (z_{dec} + 1)^3  \quad & &\quad 
n_{e^-} = n_p \quad \quad n_{\bar p} = n_{e^+} = 0\\
n_{dm} = \frac {\rho_c (\Omega_M - \Omega_{hot} - \Omega_b (1 + \frac {m_e}
{m_p}))}{m_{dm}} (z_{dec} + 1)^3 & & 
{\Omega}_{hot} = \frac {\pi^2}{30} g_* T_{cmb}^4. \label {initdens}
\eea
Knowing initial density and temperature, other quantities like chemical 
potential and distributions can be determined~\cite{earlyu}. We assume 
that the age of the Universe $\tau_0 = 14.8 Gyr$. This quantity also depends on the 
equation of state and we use it only for fixing the lifetime of WIMPZILLA. 

\subsection {Backgrounds}
Apart from CMB and relic neutrinos which are included in the initial conditions, 
we don't include any other background to high redshift distributions. For 
$z \leq 3$, we add near-IR to UV emissivity of stars~\cite{opti}~\cite{joel} 
to equation (\ref{fsol}). Far-IR which is very important for energy 
dissipation of UHECRs, 
is not added because there is very little information about its evolution with 
redshift. High energy backgrounds are not added because we want to be able to 
distinguish the contribution of remnants from other sources.\\
Adding backgrounds by "hand" evidently violates the energy conservation of 
the model, but there is not any other simple alternative method. Moreover, 
the violation is very small and comparable to numerical errors.\\
Here a comment is in order: why have we included star backgrounds and not the 
synchrotron radiation ? Star background is considered as an external source. 
By contrast, synchrotron radiation concerns high energy electrons which are 
involved in the evolution. In (\ref{bolt}), the elimination of interaction 
with an external field in a homogeneous universe means that the probability 
for production and absorption of synchrotron photons is the same. It is not 
therefore possible to consider their production without their absorption i.e. 
interaction of electrons with external magnetic field.

\section {Results}
\begin{figure}[t]
\begin{center}
\psfig{figure=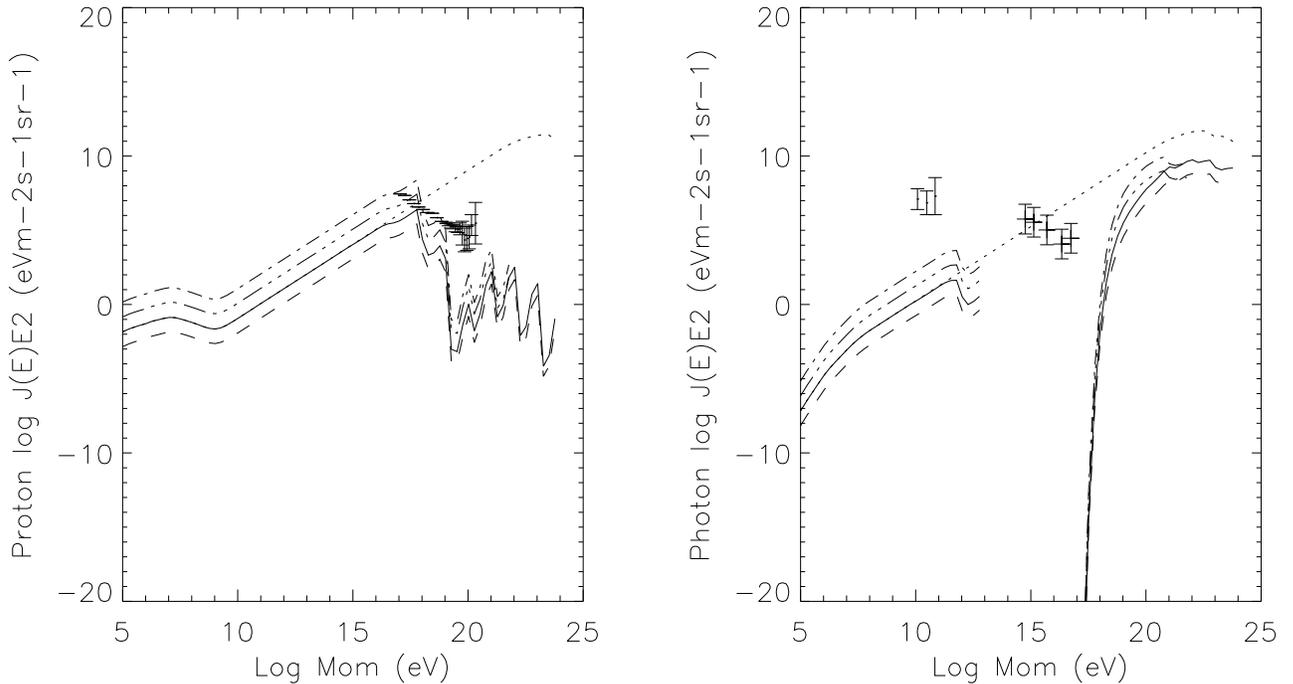,height=10cm}
\caption{Energy flux for protons and photons. Solid line 
$m_{dm} = 10^{24} eV$, $\tau = 5 \tau_0$, dot line is the spectrum 
without energy dissipation for the same mass and lifetime, dashed line 
$m_{dm} = 10^{24}eV$, $\tau = 50 \tau_0$, dash dot $m_{dm} = 10^{22} eV$, 
$\tau = 5 \tau_0$, dash dot dot dot $m_{dm} = 10^{22} eV$, 
$\tau = 50 \tau_0$. For protons, data from Air Showers detectors~\cite{crrev} 
is shown. Data for photons are EGRET whole sky background~\cite{egret} and 
upper limit from CASA-MIA~\cite{casa}.\label{fig:pgzoom}}\end{center}
\end{figure}

\begin{figure}[t]
\begin{center}
\psfig{figure=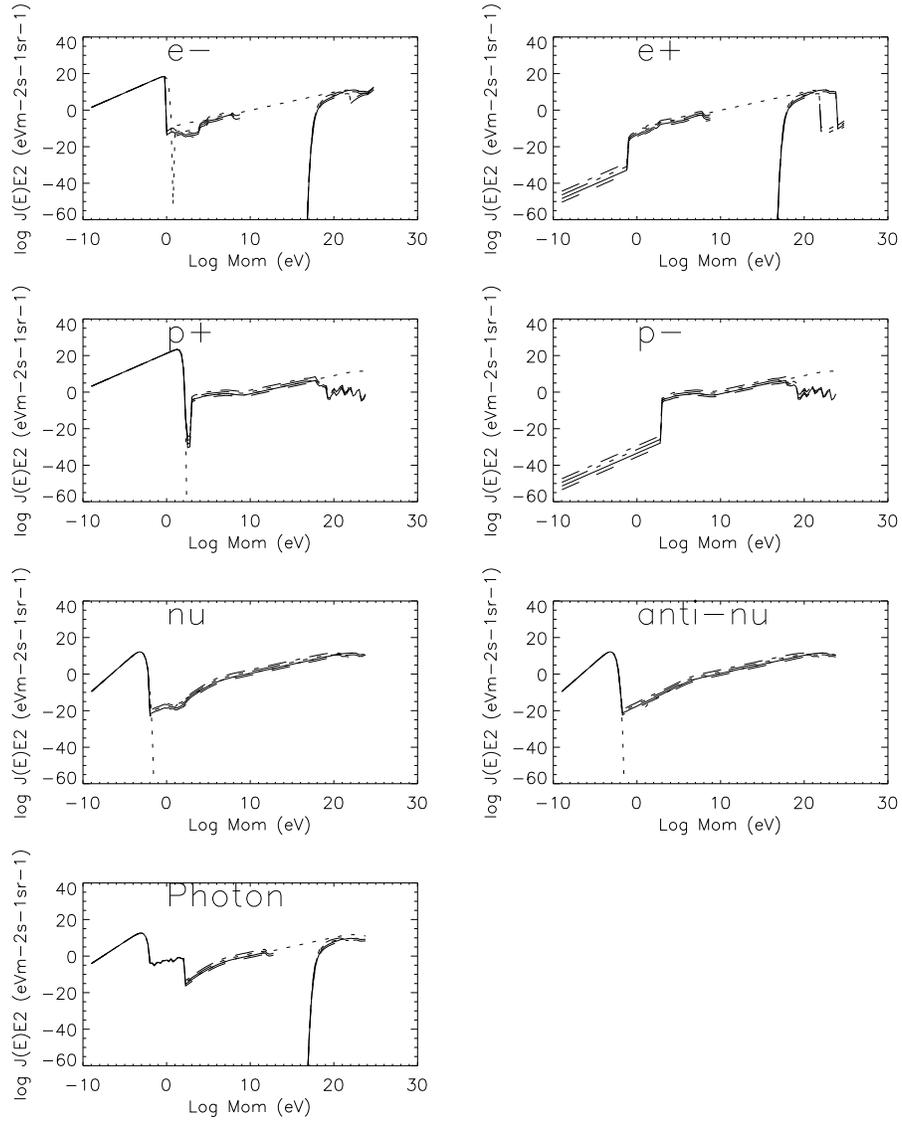,height=15cm}
\caption{Energy flux of stable species. Description of curves is the same as 
Fig.\ref{fig:pgzoom}. \label{fig:distall}}
\end{center}
\end{figure}

\begin{figure}[t]
\begin{center}
\psfig{figure=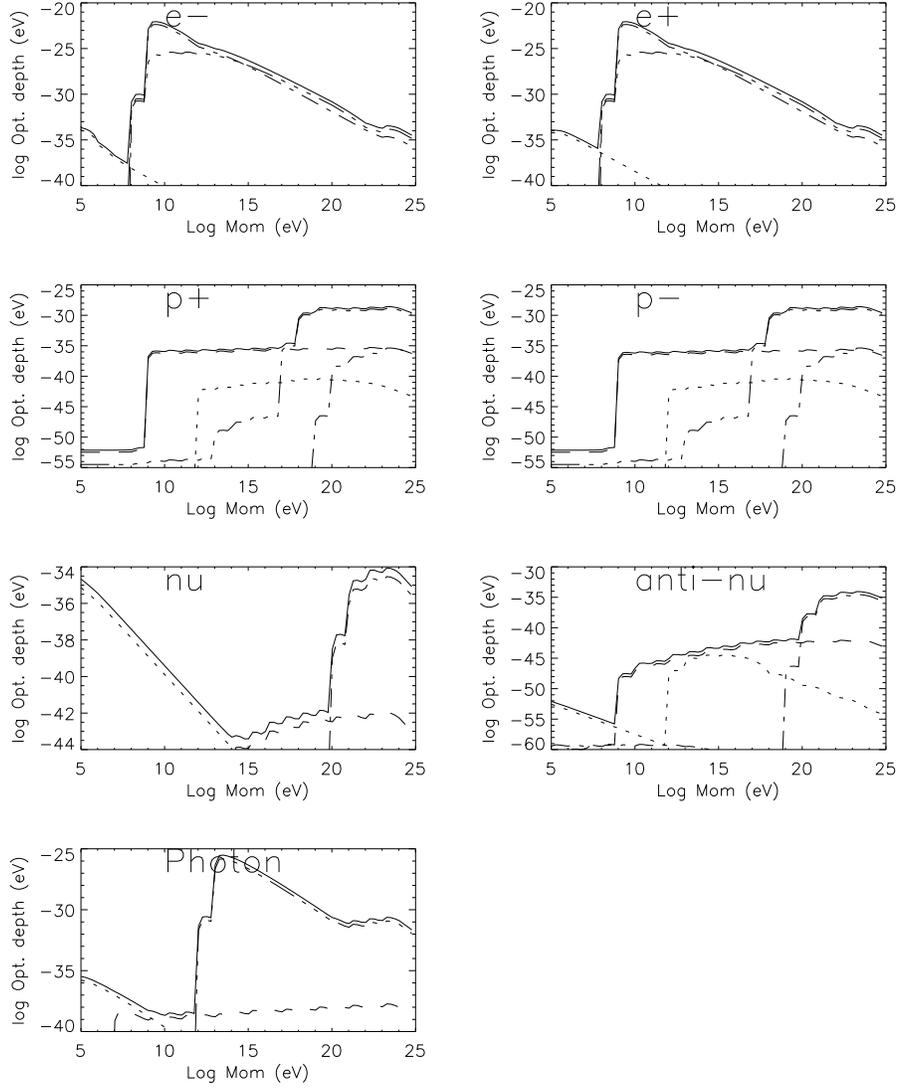,height=15cm}
\caption{Total optical depth of species and contribution of backgrounds. 
Solid line is 
total optical depth, dot line contribution of $e^\pm$, dashed line $p^\pm$, 
dash dot $\nu \& \bar \nu$, and dash dot dot dot $\gamma$. Dependence on 
lifetime and mass of UHDM is negligible.
\label{fig:optdepth}}
\end{center}
\end{figure}

Figure \ref{fig:pgzoom} shows the energy flux of high energy protons and 
photons in a homogeneous universe. Fig.\ref{fig:distall} shows the same 
quantity for all species. The GZK cutoff is very transparent. With our 
background model it begins at $E \approx 10^{18.2} eV$ for protons due to $p-\gamma$ interaction (see optical depth in Fig.\ref{fig:optdepth}) and at 
$E \sim 10^{13} eV$ for photons due to $e^\pm$ production. For purely 
kinematic reasons, proton cutoff is much shallower than photon one. The 
resonance of $p-\gamma$ interaction is very close to $p$ rest mass where 
$A(s)$ in (\ref{absint}) is very small. Moreover, division of (\ref{bolth}) by 
$E$ reduces the effect of absorption on the distribution.\\
According to Fig.\ref{fig:pgzoom}, even the shortest lifetime we have 
considered, can not explain the observed flux of protons (in contrast to 
~\cite {topdown} that assumes a 2-particle decay mode). The same figure 
shows also the flux without energy dissipation. It is compatible 
with Ref.~\cite{sarkar} which assumes a hadronic decay but does not consider 
the energy dissipation of secondary particles. In this latter case, the 
lifetime must be 
$\sim 4-6$ orders of magnitude larger than the age of the Universe. 
On the one hand, this result proves the r\^ole of a realistic model of energy 
dissipation in the estimation of mass and lifetime of UHDM. On the other 
hand, it shows the importance of clumping of Dark Matter, i.e. most of 
observed UHECRs must come from nearby sources. This conclusion is 
independent of the source of UHECRs. A similar conclusion has been obtained in 
~\cite{xx} by fitting the expected flux from extragalactic sources and from 
the Halo on the data. However they don't consider the dissipation. In the next 
section we show that even in a halo, the dissipation of photons energy is 
significant.\\
In the case of decaying UHDM hypothesis, the most important source is the 
Galactic Halo. Before trying to make a simple model of halo in the next 
section, we discuss some of other conclusions one can make from this study.\\
From Fig.\ref{fig:pgzoom} and by taking into account the fact that $p-\gamma$ 
cross-section is $\sim 10^4$ times smaller than $p-p$, with present 
statistics of UHECRs, less than one photon shower could be observed.\\
The comparison of EGRET data for $10^8 eV < E < 10^{11} eV$ with our 
calculation shows that it is compatible with a short lifetime UHDM. Future 
observations of GLAST at $E > 10^{11}eV$ is crucial for understanding the 
source of UHECRs because one expects an additional contribution from 
synchrotron radiation of ultra high energy electrons at 
$E \sim 10^{12} - 10^{14} eV$~\cite{crpro}~\cite{topdown}.\\
Fig.\ref{fig:optdepth} illustrates the total and partial optical depth, i.e. 
${\mathcal B}(t,p)$ in equation (\ref{absinth}) at $z = 0$. It shows that for 
protons with $E \sim 10^{20} eV$, even a source at $\sim 20 Mpc$ must be 
very strong to be able to provide the observed flux because its flux is 
reduced by $11$ orders of 
magnitude! The optical depth of protons must be even larger than what we have 
obtained here because we didn't take into account the far-IR and radio 
backgrounds. This 
makes difficulties for the recent suggestion by Ahn E.J., \etal~\cite{virgo} 
that Virgo cluster can be the only source of UHECRs, even if the magnetic 
field of Galactic wind is as strong as what is considered in that work. The 
main challenge is finding a conventional source with enough emissivity of CRs 
at ultra high energies. 

\subsection {CMB Distortion and Entropy Excess}
\begin{figure}[t]
\begin{center}
\psfig{figure=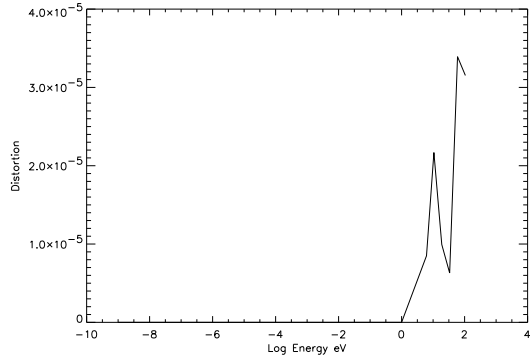,height=5cm}
\caption{Fraction of distortion in photon distribution with respect to a stable DM.\label{fig:cmbdis}}
\end{center}
\end{figure}

To see if a relatively short living UHDM can distort CMB, we reduced the 
spectrum of photons for a decaying DM from the spectrum with a stable DM. 
Fig.\ref{fig:cmbdis} shows the result for $z = 0$. There is no distortion 
at least up to $1$ to $10^8$ parts for $E \lesssim 3 eV$. However, 
one should note that this conclusion depends somehow on the cross-section 
cuts at low energies. Nevertheless, giving the fact that CMB flux at its 
maximum is $\sim 10^{20}$ times larger than the rest of the spectrum, it 
seems very unlikely 
that it can be disturbed, otherwise SZ effect due to stars had to be observed 
everywhere. The distortion of CMB anisotropy will be studied elsewhere.\\ 
We have also examined the entropy excess separately for each species. There is 
no entropy enhancement except for $e^+$ and $p^-$ which in our model are 
absent from the initial conditions. Comparing to other species, their 
contribution is very small and negligible.

\subsection {Baryon and Lepton Asymmetry Generation}
It has been suggested~\cite{houribbn}~\cite{revorg} that the decay of UHDM may 
be able to generate additional baryon and lepton asymmetry. At GUT scale, i.e. the mass scale of WIMPZILLA, we 
expect such processes and the fact that a late time decay is out of thermal 
equilibrium and satisfies Sakharov conditions for baryogenesis~\cite{earlyu} 
make growing asymmetry plausible.\\
The rate of baryonic (or leptonic) number production by decay of UHDM in comoving 
frame can be expressed as: 
\be
\frac {d(n_b - n_{\bar b})}{dt} + 3 \frac {\dot a (t)}{a (t)} (n_b - 
n_{\bar b})= \frac {\varepsilon n_{dm}}{\tau}. \label {nbaryon}
\ee
$\varepsilon$ is the total baryon number violation per decay. The solution of 
this equation is:
\bea
\Delta (n_b - n_{\bar b}) & = & \varepsilon n_{dm}(t_0) (1 - \exp (- \frac 
{t - t_0}{\tau})) \frac {(1+z_0)^3}{(1+z)^3}. \label {nbs}\\
\Delta B & \equiv & \frac {\Delta (n_b - n_{\bar b})}{2 g_* n_\gamma} = 
\frac {\varepsilon n_{dm}(t_0)}{2 g_* n_\gamma (t_0)} (1 - \exp (- \frac 
{t - t_0}{\tau})) \frac {(1+z)}{(1+z_0)}. \label {db}
\eea
If $t_0 = t_{dec}$, $\frac {n_{dm}(t_0)}{n_\gamma (t_0)} \sim 10^{-22}$ 
(for $m_{dm} = 10^{24} eV$). Therefore $\Delta B \sim 10^{-22} \varepsilon$ at 
$z = 0$. 
As $\varepsilon$ can not be larger than total multiplicity, $\sim 1000$, 
$\Delta B \lesssim 10^{-19}$, i.e. much smaller than primordial value $\sim 10^{-10}$.\\
We tested this argument by assuming $\varepsilon = 0.1$ at all energies, i.e. 
$\varepsilon_{tot} = 0.1 {\mathcal M_{tot}}$. Evidently $n_{\bar p}$ is 
smaller, but the change of $n_p$ is too small to be measured. The same is 
true for number density of leptons, but energy density of leptons with 
respect to anti-leptons increases by an amount comparable to $\varepsilon$.

\subsection {Equation of State of the Universe}
\begin{figure}[t]
\begin{center}
\psfig{figure=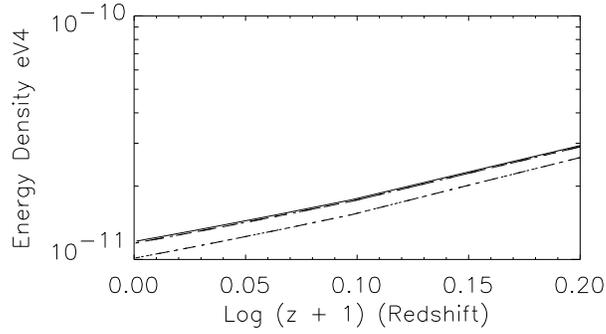,height=5cm}
\caption{Energy density of the Universe. Solid line a stable DM, dash 
line $\tau = 50 \tau_0$, dash dot $\tau = 5 \tau_0$. Dependence on the mass of 
UHDM is negligible.\label{fig:t00}}
\end{center}
\end{figure}

Decay of UHDM gradually changes part of CDM to HDM and thus changes the 
equation of state of the Universe. Fig.\ref{fig:t00} shows the variation of 
equation of state. For $\tau \sim 50 \tau_0$ or larger, it would be too small 
to be measurable. For smaller $\tau$, DM decay plays the r\^ole of a running 
cosmological constant. A complete study of this issue and comparison with data 
is under preparation.

\section {Halo}
To see the effect of clumping of a decaying UHDM on the flux of UHECRs, here 
we try to make a very simple model. A complete treatment of halos will be 
reported elsewhere.\\
We consider a halo as a uniform over-density with a limited size at $z = 0$. 
This is simulated by following the decay of UHDM and evolution of remnants 
for a time comparable to the propagation time in the halo. 
Evolution equation and energy binning is taken to be the same as in the 
homogeneous universe case with $a(t) = cte$. Because we want to study the 
propagation of remnants in a volume comparable to the Galactic Halo, we 
consider time steps equivalent to $10 kpc$. Our tests show that after a few 
steps ($\sim 7$), the accumulation rate of ultra high energy particles becomes 
very slow. We consider two cases. In the first case we simply evolve 
distributions for a number of steps (up to 30). In the second case, after some 
steps, we stop the decay of the Dark Matter to simulate an inner halo of 
MACHOs. Then, the evolution is continued for more $5$ steps (i.e. 
$50 kpc$)~\cite{machobs} to simulate propagation through MACHOs. Evidently this model is very approximative. We use it only to make a crude 
estimate of production and absorption of UHECRs.

\subsection {Initial Conditions and Galactic Backgrounds}
Galactic baryonic matter is taken to have a thermal distribution with 
$T_b = 10^4 K$. We assume that baryonic over-density is biased with 
respect to DM i.e. the fraction of baryons to DM is larger than its mean 
value in the Universe. With these assumptions the initial number densities 
can be expressed as:
\bea
n_p = \frac {b \delta \rho_c}{m_p} \quad \quad n_{e^-} = n_p  & & \nonumber\\
n_{\bar p} = n_{e^+} = 0 & & \nonumber\\
n_{dm} = \frac {(1 - b) \delta \rho_c}{m_{dm}} \label {haloinit}
\eea
$b$ is the fraction of baryons in the halo. In the following $b = 0.3$, i.e. 
$\sim 2$ times primordial value~\cite{bias} in the cosmological model 
explained above. $\delta$ is the mean over-density of the Halo. Inspired 
by universal halo density distribution of NFW~\cite{nfw}, we consider a halo 
with characteristic radius (i.e. virial radius) $r_{200} = 0.12 Mpc$. 
According to NFW distribution and by definition this means $\delta = 200$. 
These parameters defines a halo of mass $M_{H} = 6 \times 10^{12} M_\odot$.\\ 
Neutrino density is assumed to be the same as relic at $z = 0$. For photons, 
in addition to CMB, we consider a galactic background as the following:\\ 
Galactic IR and visible backgrounds are not very well known. We use the 
results of the model developed by DIRBE group for detection of 
extragalactic component of the IRB~\cite{dirbe}. We consider the observed 
value of IRB after elimination of Inter-Planetary Dust (IDP) contribution as 
the galactic background. It is just an estimation of average galactic IRB. It 
is not clear if we can extend the local value of IRB to whole galaxy or take 
it as a representative average. For this reason we also increase it $10$ times 
(probably an extremely high value) to see the effect on the energy dissipation 
of UHECRs (see below for conclusions). Our simulation does not include radio 
background.\\
For soft and hard X-Ray galactic backgrounds, we use the model developed for 
extraction of extragalactic component from ROSAT and ASCA 
observations~\cite{galx}. It considers GXB as two thermal components, a soft 
component with $T_{sx} = 70 eV$ from Local Bubble, and a hard component with 
$T_{hx} = 145 eV$ from hot gas, probably in the Halo. We add also the 
extragalactic component for $0.25 keV < E < 10 keV$. 

\section {Results}
\begin{figure}[t]
\begin{center}
\psfig{figure=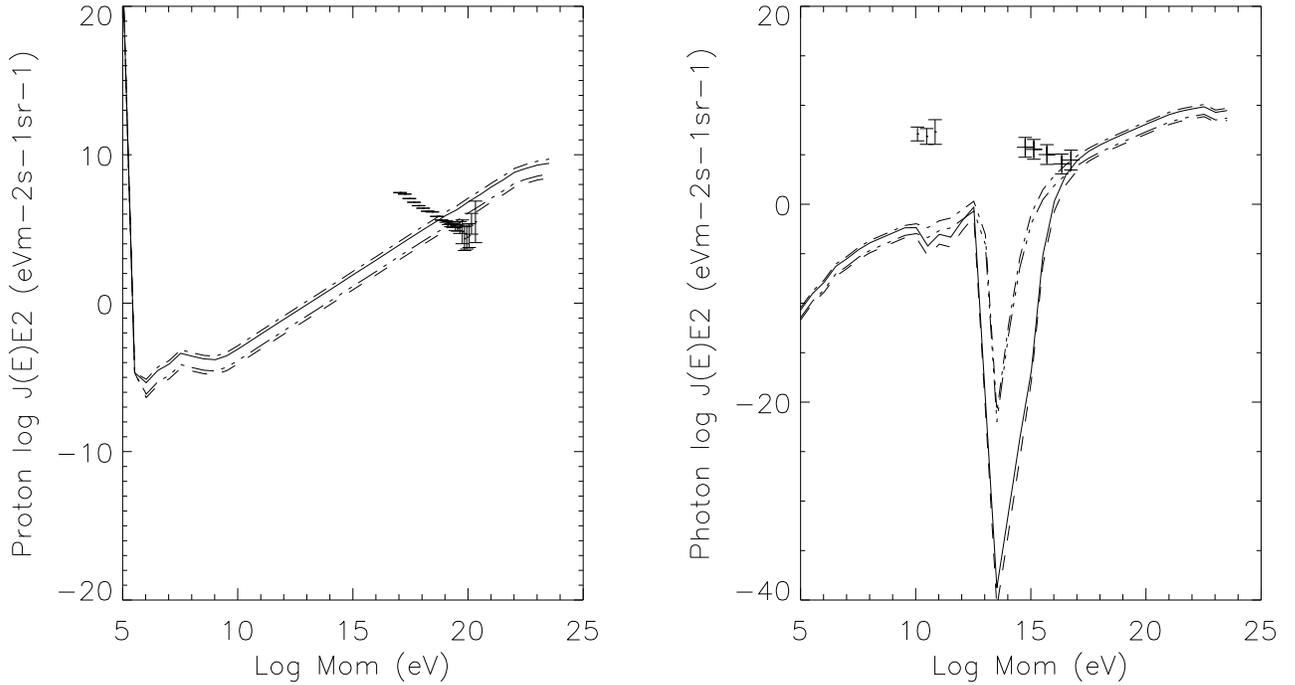,height=10cm}
\caption{Flux of high energy protons and photons in a uniform clump. $m_{dm} 
= 10^{24} eV$, $\tau = 5 \tau_0$ and $\tau = 50 \tau_0$. Dash dot and dash dot dot dot lines presents UHDM halo. Solid and dashed lines show a halo of UHDM 
and MACHOs. Data is the same as in Fig.\ref{fig:pgzoom}. For protons the 
effect of increasing lifetime of UHDM is more important than presence of 
MACHOs. Photons trough is more sensitive to presence of MACHOs.
\label{fig:machhalo}}
\end{center}
\end{figure}

Fig.\ref{fig:machhalo} shows the distribution of high energy protons and 
photons for a uniform halo and for a halo that its inner part is composed of 
MACHOs. Only the result for $m_{dm} = 10^{24} eV$ with $\tau = 5 \tau_0$ and 
$\tau = 50 \tau_0$ is shown. Because of importance of the spectrum close to 
observed trough in the interpretation of the results, Table \ref{tab:trough} 
summarizes the numerical value of simulated and observed spectrum.\\
For photons the trough of GZK cutoff is shallower 
than in a homogeneous universe and for protons it is practically absent. 
Consequently, the calculated flux at $E \sim 10^{19.5} eV$ is somehow higher 
than observation. However, at higher energies simulation results specially 
one with a MACHO halo is in the $1-\sigma$ error range of the observations.\\ 
The simplest explanation of having a smaller observed flux close to the 
minimum of the spectrum can be the need for increasing the lifetime somehow 
(but not by many orders of magnitude as suggested by previous 
works~\cite {sarkar}~\cite {topdown}). However, the lack of a minimum in the 
simulated spectrum\footnote {In fact 
optical depth has a maximum at $E \approx 3.3 \times 10^{19} eV$.} means 
that in some way our simulation is not exact. It can be due to a too simple 
halo model. Other possibilities are a different (probably harder) decay 
spectrum and/or energy loss in the Galactic magnetic field. If these 
suggestions are true, fluxes will be smaller and there is no need for 
increasing the lifetime. A more complete simulation 
of the Halo and magnetic field and a better understanding of non-baryonic DM 
distribution is necessary for making any definitive conclusion.\\
The minimum in the spectrum can also be interpreted as a wider 
distribution of sources. In this case, it is hardly probable that sources 
responsible for CRs at lower 
energies can explain this behavior because even if the spectrum of these 
sources can be extrapolated with the same slop to higher energies, it can 
not explain the rising slop of the spectrum.\\
As the IR background is crucial for energy dissipation of 
UHECRs, we have also increased it to $10$ times of the DIRBE model to see 
the effect. Proton flux at high energies slightly decreases, but it can not 
explain the observed minimum. We have also tested the distortion of the CMB by 
remnants as described for a homogeneous universe. There is no distortion up 
to at least $1$ to $10^8$ for $E \lesssim 10 eV$.\\
Summarizing this section, it seems that a decaying UHDM with a lifetime as 
short as $\tau \sim 10 - 100 \tau_0$ is not ruled out by present 
observations.
\begin{table}[t]
\caption{Energy flux of UHECRs close to GZK cutoff. Note that there is not a 
complete agreement between estimated flux by different Air Shower detectors. 
The values presented here are mostly based on AGASA data before 1998.\label{tab:trough}}
\vspace{0.2cm}
\begin{center}
\footnotesize
\begin{tabular}{|c|c|c|c|}
\hline
\raisebox{0pt}[13pt][7pt]{$\log (E) eV$} &
\raisebox{0pt}[13pt][7pt]{$\tau = 5 \tau_0$, $\log E^2 J (E)$} &
\raisebox{0pt}[13pt][7pt]{$\tau = 50 \tau_0$, $\log E^2 J (E)$} &
\raisebox{0pt}[13pt][7pt]{Observed $\log E^2 J (E)$} \\
\raisebox{0pt}[13pt][7pt]{} &
\raisebox{0pt}[13pt][7pt]{$m^{-2} \sec^{-1} sr^{-1}$} &
\raisebox{0pt}[13pt][7pt]{$m^{-2} \sec^{-1} sr^{-1}$} &
\raisebox{0pt}[13pt][7pt]{$m^{-2} \sec^{-1} sr^{-1}$} \\
\hline
\raisebox{0pt}[12pt][6pt]{$18.525$} & 
\raisebox{0pt}[12pt][6pt]{$5.4$} & 
\raisebox{0pt}[12pt][6pt]{$4.4$} & 
\raisebox{0pt}[12pt][6pt]{$6.15 \pm 0.01$}\\
\hline
\raisebox{0pt}[12pt][6pt]{$19.025$} & 
\raisebox{0pt}[12pt][6pt]{$5.9$} &
\raisebox{0pt}[12pt][6pt]{$4.9$} &  
\raisebox{0pt}[12pt][6pt]{$5.5 \pm 0.05$}\\
\hline
\raisebox{0pt}[12pt][6pt]{$19.525$} & 
\raisebox{0pt}[12pt][6pt]{$6.35$} & 
\raisebox{0pt}[12pt][6pt]{$5.3$} & 
\raisebox{0pt}[12pt][6pt]{$5.08 \pm 0.2$}\\
\hline
\raisebox{0pt}[12pt][6pt]{$20.1$} & 
\raisebox{0pt}[12pt][6pt]{$7$} & 
\raisebox{0pt}[12pt][6pt]{$6$} & 
\raisebox{0pt}[12pt][6pt]{$5.4 \pm 1.4$}\\
\hline
\raisebox{0pt}[12pt][6pt]{$20.525$} & 
\raisebox{0pt}[12pt][6pt]{$7.36$} & 
\raisebox{0pt}[12pt][6pt]{$6.3$} & 
\raisebox{0pt}[12pt][6pt]{$\sim 5 \pm 2$}\\
\hline
\raisebox{0pt}[12pt][6pt]{$21.025$} & 
\raisebox{0pt}[12pt][6pt]{$8.09$} & 
\raisebox{0pt}[12pt][6pt]{$6.8$} & 
\raisebox{0pt}[12pt][6pt]{$$}\\
\hline
\end{tabular}
\end{center}
\end{table}

\section {Conclusions}
The main purpose of this work was showing the importance of a realistic 
physical model for finding the answer to the mystery of UHECRs and 
introducing readers to the program that has been developed for achieving this 
goal.\\
We showed that ${\mathcal R} \equiv \xi {\tau}_0 / \tau \sim 0.1-0.01$ 
where $\xi$ is the contribution of UHDM in the Dark Matter. This value is 
larger (or equivalently the lifetime is shorter for the same contribution) 
than what has been suggested in previous works which don't 
consider the dissipation~\cite{xx}~\cite{sarkar}. If a more realistic halo 
model confirms this conclusion, the cosmological implication of a UHDM can 
be important.\\
We have studied a special decay mode. Any other mode that produces 
{\it invisible} WIMPs or leptonic or semi-leptonic modes decreases the 
lifetime of UHDM. If more nucleon are produced, the lifetime must be longer 
but it seems less probable than other cases.\\
We showed the reciprocal influence of UHECRs and backgrounds on each others. 
Consequently, whatever the source of UHECRs, it is extremely important to 
correlate their observations to the observation of high energy photon and 
neutrino backgrounds.\\
This work is the first step to a comprehensive study of the effects of a 
decaying 
Dark Matter. Other issues like a realistic model for halos, effects on the 
determination of cosmological parameters and equation of state and comparison 
with more data from cosmic rays and high energy backgrounds remain for future 
works.\\
{\bf Note:} Shortly after completion of this work the Lake Baikal 
Experiment collaboration~\cite {baikal} have published their upper limit on 
the flux of high energy neutrinos. It is well above what is obtained in our 
simulation (see Fig.\ref {fig:optdepth}).

\section* {Appendix 1: Fragmentation in MLLA}
\begin{figure}[t]
\begin{center}
\psfig{figure=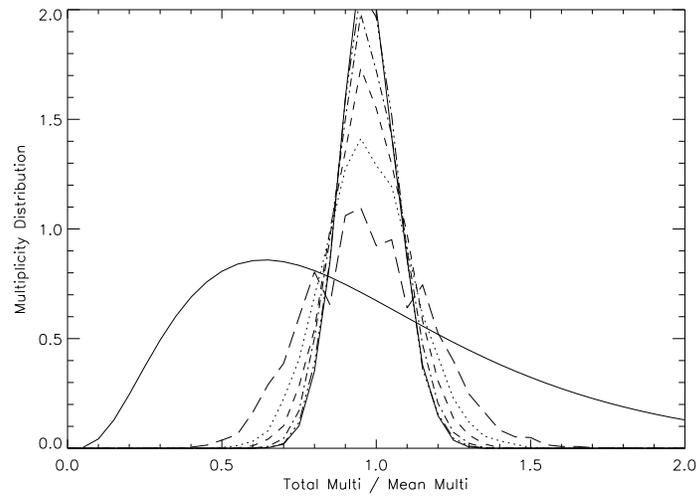,height=7cm}
\caption{Distribution of multiplicity in hadronization of a pair of gluon jets.
Solid line is the highest energy $E_{CM} = 10^{20} eV$; Long dash is lowest 
energy $E_{CM} = 10^{11} eV$. KNO distribution is shown also (solid line). 
\label{fig:multdist}}
\end{center}
\end{figure}

The MLLA treats fragmentation as a Markov process. Consequently, the 
differential 
multiplicity is proportional to splitting function $P(x)$(See e.g. ~\cite{lphd}):
\be
\frac {d{\mathcal M} (x_E)}{dx_E} \propto P (x) \quad \quad x_E = \frac {E}{E_j}. 
\label {diffmult}
\ee
For a gluon jet and at $x_E \ll 1$, $P (x_E) \sim \frac {1}{x_E}$, and one 
expects that:
\be
\frac {d{\mathcal M} (x_E)}{d(ln (x_E))} \propto \mbox {Total Num. of splitting} 
\propto E_j. \label {lnx}
\ee
In PYTHIA however, at high energies $E \gtrsim 100 GeV$, increasing CM energy 
just increases the probability of having more high energy fragments which 
simply escape fragmentation. A consequence of this behavior is the narrowing 
of the multiplicity distribution (See Fig.\ref {fig:multdist}) in contrast to 
theoretical prediction i.e. KNO~\cite {kno} scaling for $E_j \rightarrow 
\infty$. This limit is obtained at parton level. However, if LPHD is valid, 
one 
expects the same type of behavior at hadron level. Another factor that can 
explain, at least partially, the deviation from theory is the decay of hadrons 
in our simulation. It increases total multiplicity more than its statistical 
variation and makes the distribution narrower, but it can not explain the 
absence of a tail of large multiplicity events. Even after decay, these events 
should keep their difference with average. As both MLLA and Monte 
Carlos fail to reproduce observations in relatively low 
energies~\cite{qgdiff}~\cite{fragrev}, the exact behavior at high energies is 
not clear.

\section* {Appendix 2: Production Integral for $2 \rightarrow 2$ Processes}
Cross-sections are Lorantz invariant. However, to have a unique expression for 
what are simulated and what are calculated analytically, we determine all of 
them in their CM.\\
The triple integral in (\ref{prointh}) depends on three momentum variables 
corresponding to the momentum of two incoming particles and one of the 
outgoing particles that its evolution is under calculation. In the case of a 
$2 \rightarrow 2$ process, the 
amplitude of the momentum of final particles depends only on s. If:
\be
{\bf p} = (E, p \cos \phi \sin \theta, 
p \sin \phi \sin \theta, p \cos \theta). \label {pll}
\ee
is the 4-momentum of outgoing particle, and ${\bf p'}$ is its counterpart in 
the CM:
\bea
{{\bf p'}}^2 & = & \frac {(s - m^2 - m'^2)^2 - 4 m^2 m'^2}{4s}. \label {pcm}\\
{\bf p'} & = & \Gamma {\bf p}
\eea
$m$ and $m'$ are the mass of out-going particles, $\Gamma$ is 
the boost matrix. The equality of two expressions for ${\bf p'}$ leads 
to an equation that can be solved for one of the angular variables in 
(\ref{pll}). The calculation is tedious but strait forward. With respect to 
$\phi$, the equation is $4^{th}$ order but in the case of a homogeneous 
cosmology where the boost matrix depends only on the relative angle between 
incoming particles, it depends only on $\tan^2 (\frac {\phi}{2})$ and is 
analytically solvable. In this way the integration over $\phi$ in 
(\ref{prointh}) reduces to sum of integrand evaluated at the roots of the 
equation.\\
This method provides a general way to deal with this problem and is more 
convenient than doing calculation for each cross-section separately.\\

{\bf Acknowledgement} I would like to thanks people of Strasburg Observatory 
for their kindness.

\end{spacing}

\pagebreak

\pagebreak

\pagebreak

\pagebreak

\pagebreak

\pagebreak

\pagebreak

\pagebreak

\pagebreak

\end {document}